\documentclass[a4paper]{jpconf}
\usepackage{graphicx}
\begin{document}
\title{BEACH 2014 Theory Summary}

\author{Sebastian J\"ager}

\address{Department of Physics and Astronomy, University of Sussex,
  Falmer, Brighton BN1 9QH, UK}

\begin{abstract}
I summarize key aspects of the quest for physics beyond the Standard Model
in flavour physics as discussed at the BEACH 2014 conference in
Birmingham. 
\end{abstract}

\section{Introduction}
\begin{figure}[h]
\begin{center}
\includegraphics[scale=0.3,trim=5cm 0 0 0]{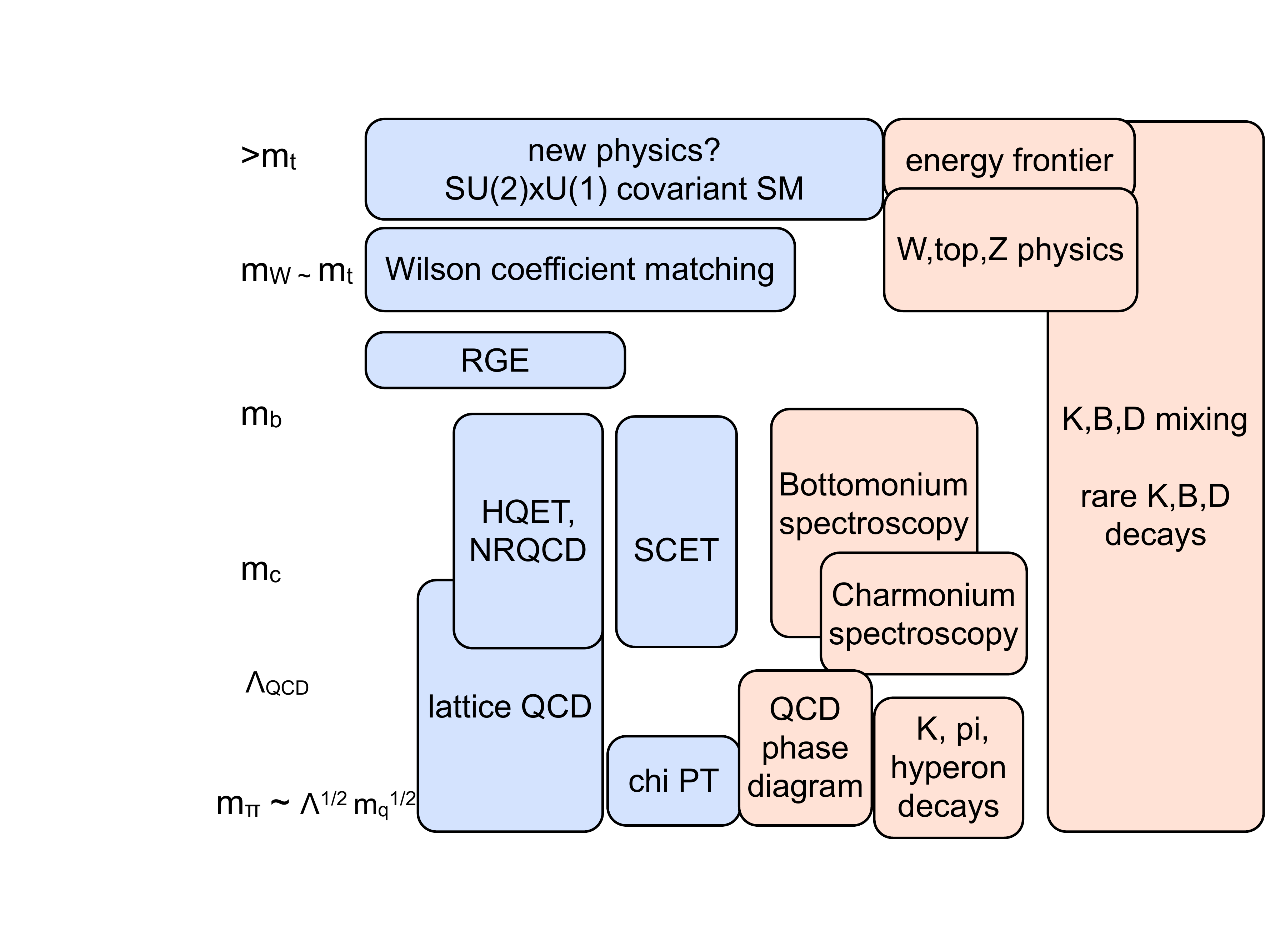}
\end{center}
\caption{\label{fig:scales}Energy scales, observables, and techniques relevant
  to flavour physics.}
\end{figure}

I thank the organisers for inviting me to BEACH 2014 in Birmingham
and bestowing on
me the honour of giving the theory summary. The many excellent
theoretical talks, many of them review talks in their own right, make
both for a rich source of topics and a challenge of selection; I
apologise at the outset for omissions. The conference started with the
low-energy QCD effects in spectroscopy and Kaon physics and ended with
the early universe, reflecting appropriately the wide range of energy
scales relevant to, and probed through, flavour physics (Figure \ref{fig:scales}).
As indicated in the figure, flavour and heavy-quark physics has seen
the birth and use of many fruitful theoretical concepts, including the weak
Hamiltonian and its renormalisation-group evolution,
heavy-quark expansions/HQET/NRQCD and soft-collinear effective theory, and
provides the impetus for many developments in lattice QCD.

\section{Flavour physics and the BSM landscape}
Flavour physics played a key role in constructing the Standard Model
(SM), including the invention of
 Cabibbo mixing to address a $2\sigma$ tension in
weak decay data \cite{GellMann:1960np}, the resolution of a naturalness problem in
the $K_L - K_S$ mass difference based on the then hypothetical charm
quark \cite{Gaillard:1974hs},
and the prediction of a third generation
by Kobayashi and Maskawa \cite{Kobayashi:1973fv} to accommodate  CP
violation in $K_L$ decay ($\epsilon_K$).

With the discovery of a scalar particle consistent, so far, with
the SM Higgs boson, another
naturalness problem looms larger than ever in the hierarchy between the
electroweak scale and any fundamental scale at which new degrees of
freedom appear, including $M_{\rm Planck}$, $M_{\rm GUT}$, $M_{\rm
  seesaw}$. The known natural beyond-SM (BSM) scenarios all involve, so far
hypothetical, particles at the TeV scale and were
reviewed in the opening talk by Kamenik \cite{talk-Kamenik}. They all bring in
new sources of flavour violation. Will flavour physics again be the
lead in constructing the next Standard Model? Nobody knows, but it
is entirely possible. Indeed, there are several
interesting puzzles in the data, some new and some older but
persistent, that have been discussed at this
meeting. No BSM particles have
been identified in the high-$p_T$ experiments yet; this certainly
raises the importance of virtual probes of the kind discussed at
this conference. 
Even without a BSM discovery so far, the wealth of data has led to a
great deal of bottom-up phenomenology in the last few years, largely
phrased in terms of ad-hoc models or fits to effective field theory
parameters, at the expense of top-down model building; indeed there
was not much discussion of model-specifics at this meeting.

The discovery power of flavour physics arises as follows.
A generic FCNC amplitude behaves as
\begin{equation}
  {\cal A} = {\cal A}_{\rm SM} + {\cal A}_{\rm BSM} .
\end{equation}
A typical observable, say a rare decay rate, provides a (schematic) constraint
\begin{equation}     \label{eq:constrdec}
    |{\cal A}_{\rm BSM}|^2 + 2\,{\rm Re}\,{\cal A}_{\rm
               BSM}^* {\cal A}_{\rm SM} 
 \;     = \; \Gamma_{\rm exp} (1 \pm \Delta^{(\rm exp)})
            - |{\cal A}_{\rm SM}|^2 (1 \pm \Delta^{(\rm SM)}) .
\end{equation}
Hence, although new physics decouples as $M_Z^2/M_{\rm NP}^2$, flavour
observables can probe well beyond the energy frontier provided sufficient
statistics and theoretical precision are available. It is perhaps worth
comparing the situation in flavour physics with that in precision Higgs
physics. Assuming a ``little hierarchy'' $M_Z \ll M_{\rm NP}$, the
leading BSM physics can be parameterized in terms of a large number
of dimension-six, $SU(2)_W \times U(1)_Y$ invariant operators
\cite{Buchmuller:1985jz,Grzadkowski:2010es}. $B$ decays alone
probe more than 100 operators even when assuming lepton flavour conservation,
far more than in the Higgs case.
(Disentangling them is a formidable but not impossible task, see below.)

Figure \ref{fig:df2bounds} \cite{talk-Kamenik} shows the energy scales
probed by the four types of meson-antimeson mixing.
\begin{figure}[h]
\begin{center}
\includegraphics[scale=0.3,trim=2cm 9cm 15cm 1cm]{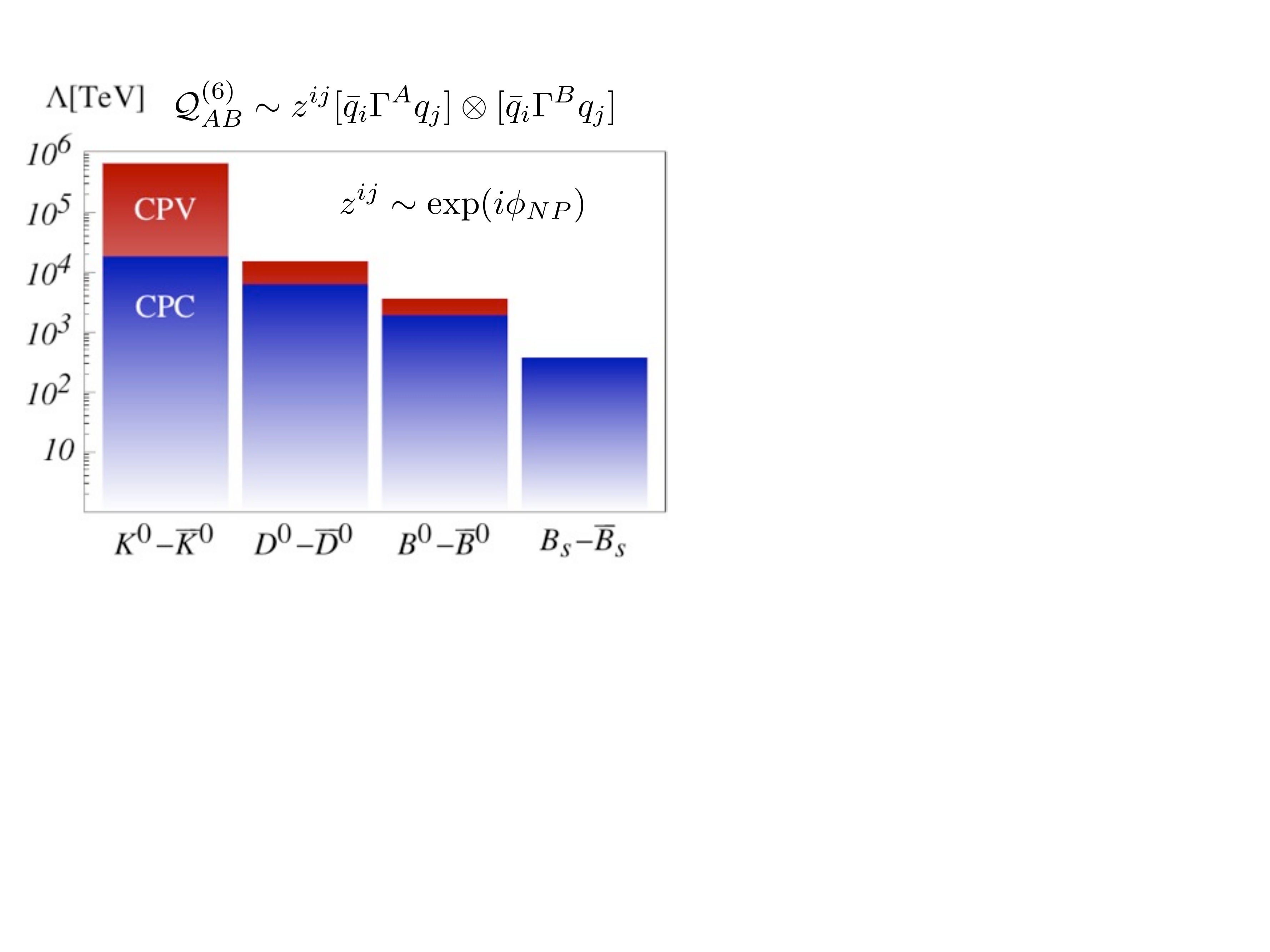}
\end{center}
\caption{\label{fig:df2bounds} Generic bounds on the scale of CP violating and
  CP-conserving new physics from meson-antimeson mixing as shown by
  Kamenik \cite{talk-Kamenik}.}
\end{figure}
 The most stringent constraint
comes from $\epsilon_K$, pointing either to
(approximately) CP-conserving new physics, to a
substantial ``little hierarchy'' between the weak scale and the new
physics, or to some kind of flavour suppression mechanism resulting in
small mixing angles suppressing strangeness-changing neutral-current transitions.
CP-conserving observables provide a
less stringent constraint, especially for Kaons, but still apply in
the absence of new sources of CP violation. It is worth noting that
this situation can change qualitatively in the future, in
particular with progress in lattice QCD that seems imminent (see
below). Once a deviation is seen in one or more experiments, flavour
physics may guide us toward an understanding of the origin of flavour,
perhaps in the form of a dynamical theory, perhaps based on symmetry
principles. 

From the phenomenological point of view, matching the experimental
precision comprises
disentangling  perturbative short-distance physics from long-distance
nonperturbative QCD effects, and calculating the latter or determining
them from data was the subject of a number of presentrations at this
conference.

\section{Kaons}
CP violation in $K-\bar K$ mixing ($\epsilon_K$) is the prototypical
flavour precision observable and generically provides the most
stringent constraints on physics beyond the SM \cite{talk-Kamenik}.
It originates from $\Delta F=2$ box diagrams with $W$ and $u$, $c$,
$t$ quarks. The dominant contribution, involving internal top quarks,
needs to be complemented by lattice calculations of the
nonperturbative normalisation $B_K$, which is given in terms of
the matrix element of a local
operator,
$$
B_K \propto \langle \bar K^0 | (\bar s_L \gamma^\mu d_L) (\bar s_L \gamma_\mu
d_L) | K^0 \rangle .
$$
Including
perturbative calculations of short-distance charm and up
contributions, one has \cite{Brod:2011ty}
\begin{equation}   \label{eq:ds2pred}
   |\epsilon_K| = (1.81\pm 0.28) \times 10^{-3}, \qquad
   \Delta M_K^{\rm SD} = (3.1 \pm 1.2) \times 10^{-15} \;\mbox{GeV} ,
\end{equation}
while experiment gives \cite{Agashe:2014kda}
\begin{equation}  \label{eq:ds2exp}
   |\epsilon_K|^{\rm exp} = (2.228\pm0.011) \times 10^{-3}, \qquad 
   \Delta M_K^{\rm exp} = (3.484 \pm 0.006) \times 10^{-15} \;\mbox{GeV} .
\end{equation}
Comparing the measured value of $\epsilon_K$ to the theory prediction
constrains the CKM unitarity triangle and allows to put constraints on
BSM physics, which can involve additional local operators. The
constraint from $\Delta M_K$ is less stringent due to the larger error
and the unknown long-distance contribution, which could interfere destructively with a
possible BSM contribution.

In fact, as discussed by Christ \cite{talk-Christ}, already for $\epsilon_K$
accuracies have now reached  the level where both refined
lattice renormalisation schemes (or perhaps a direct
calculation of the RG-invariant bag parameter $\hat B_K$) and precision
determinations of the charm and up loop contributions become
important. The latter two correspond to non-local matrix elements
of the sort
$$
\langle \bar K^0 | T \left\{ [(\bar s_L \gamma^\mu d_L) (\bar c_L \gamma_\mu c_L)](x)
     [(\bar s_L \gamma^\mu d_L) (\bar c_L \gamma_\mu c_L)](0) \right\} | K^0 \rangle .
$$
These include a perturbative part from scales $\sim m_c$
(although $\alpha_s(m_c)$ is relatively
close to the strong-coupling regime), which
matches onto the same local operator, and further contributions from scales
$\sim \Lambda_{\rm QCD}$.  Lattice QCD
calculations with dynamical charm quarks are starting to access the required
correlation functions directly.
These developments also raise the exciting prospect of extending the energy reach of 
 $\Delta M_K$ as a probe of CP-{\em conserving} new physics by an
 order of magnitude in the near future.

Beyond mixing, the super-rare decays $BR(K^+ \to \pi^+ \nu \bar \nu)$
and $BR(K_L \to \pi^0 \nu \bar \nu)$ provide two further very clean probes of physics beyond
the Standard Model, with SM theory errors at the 10\% level. Though
not discussed in any detail from a theory point of view at this conference,
experiment is on the move: The goal at NA62 ($K^+$, CERN, now
running), where the Birmingham group is heavily involved,
is a 10\% measurement (assuming the SM branching ratio), while
KOTO (J-PARC, restarting in 2015) expects to observe the $K_L$ mode, which violates CP.
Both will provide stringent constraints on, or measurements of, the $sdZ$ vertices,
very sensitive to $SU(2)$-breaking BSM effects such as stop-scharm mixing originating
from flavour-violating $A$-terms (eg \cite{Colangelo:1998pm,Buras:2004qb,Isidori:2006qy}).
Exciting lattice prospects were discussed \cite{talk-Christ}  for a holy grail of Kaon theory,
the two two-body $K \to \pi \pi$ decay amplitudes ($I=0$ and $I=2$).
From a BSM point of view, the experimental value of the direct-CP violation observable
$\epsilon'/\epsilon = (1.66 \pm 0.23) \times 10^{-3}$
\cite{Agashe:2014kda}
is complementary to mixing and rare $K$ decay in that it conveys
information about a
variety of short-distance couplings,
including the $sdZ$ vertex and the chromo-dipole couplings $\bar s
\sigma^{\mu\nu} P_{L,R} d G_{\mu\nu}$. From an SM/QCD point of view, there
is also the question why the decay amplitude into the $I=0$ final state is more
than 20 times larger than that into the $I=2$ state (the $\Delta
I=1/2$ rule), which predates the advent of the SM. Both questions require the computation
of the complex amplitude ratio $A(I=2)/A(I=0)$. There is
a long history of sophisticated and rather successful calculations based on
chiral perturbation theory and the large-$N_c$ limit, as reviewed by Pich
\cite{talk-Pich}. The full-QCD, $N_c=3$ results
have been elusive so far. It is all the more remarkable that the RBC/UKQCD
collaboration now appears on track for a fully realistic QCD
calculation, with the prospect of a 20\% calculation of
$\epsilon'/\epsilon$. There are many more uses of Kaons, such as
constraining the CKM matrix element $V_{us}$, for which I refer to
\cite{talk-Christ,talk-Pich}.

\section{Heavy quarks}
The three heavy quarks benefit, to varying degree, from a hierarchy $m
\gg \Lambda$, which gives rise to heavy-quark spin symmetries
and is the basis for various effective field theories.
Brambilla reviewed the theory and applications of quarkonia spectra
and furthermore discussed the nature of heavy-quark bound
states close to or above threshold, where,
unlike for states well below threshold, the heavy-quark expansion
(formalised through NRQCD and pNRQCD) no longer leads to a calculable
framework in terms of potentials (which could be determined
perturbatively or via lattice QCD simulations) \cite{talk-Brambilla}. 
These include in particular states such as the $X(3872)$ that appear
to behave as a molecular state made of two heavy-light systems. The
nature of the $X(3872)$, which within experimental accuracies sits
right on top of the $D^{0*} \bar D^0$ threshold, was discussed in more
detail by Nieves \cite{talk-Nieves}. If this picture is correct, it
predicts a variety of other nearby states, and we may hope that
more data will shed light on this very difficult multi-scale QCD problem.
The fate of charmed hadrons in the quark gluon plasma, and other
current topics in high-temperature QCD, were reviewed by
Steinheimer \cite{talk-Steinheimer}. Heavy and multiply-heavy
baryons were the subject of  talks by Azizi \cite{talk-Azizi} and
by Erkol \cite{talk-Erkol}.

\section{Charm}
Charm mixing and decay provide sensitivity to short-distance physics
similarly to the Kaon case. Unfortunately, the benefits of having a
(slightly) heavy quark are outweighed by the loss of chiral
perturbation theory together with an adverse CKM/GIM
structure, which leads, in the SM, to a complete long-distance dominance
for mixing as well as decays, including rare ones. 
Petrov discussed $D -\bar D$ mixing and CP violation in hadronic charm
decays \cite{talk-Petrov}. If one assumes the absence of
large cancellations between the SM and new-physics contributions, the
measured mass difference bounds the new physics scale, but less so
than in the Kaon case (Figure \ref{fig:df2bounds}).
The non-observation of CP violation in mixing gives a
(stronger) constraint  (on CP-violating new physics); in this case the
long-distance dominance in the SM leads to a suppression: the
three-generation, CP-violating SM is approximately reduced to a
two-generation, CP-conserving theory.

The situation is quite different for the difference of direct CP asymmetries,
$\Delta a_{\rm CP} = a_{\rm CP,KK} - a_{\rm   CP,\pi\pi}$. Previously
measured to be $3\sigma$ away from zero to much excitement, $\Delta
a_{\rm CP}$
is no longer significant as of 2013. In order to turn this into a
constraint on the new physics scale, one would require knowledge of (or make
estimates about) the strong phases of the SM contributions.

Fajfer discussed a large number of rare semileptonic and radiative $D$
decay rates and asymmetries \cite{talk-Fajfer}. All of them suffer
from large and rather uncertain, long-distance-dominated SM
contributions, but there are a few examples
where the experimental bound still significantly exceeds the SM
estimates and new physics could give a dominant contributions, such as
$D^0 \to \mu^+ \mu^-$.

Overall, the power of charm to detect new physics still appears relatively bleak, with
the exception of mixing. Eventually, lattice simulations with
dynamical charm quarks may open up charm to precision phenomenology,
at least for the $\Delta C=2$ case (and perhaps beyond \cite{talk-Petrov}).

\section{Beauty and Truth}
$B$ physics comprises the lion's share of observables in flavour
physics, many of them sensitive to SM parameters and/or new
physics. The advent of the LHCb experiment has brought a wealth of
measurements of new decay  modes and unprecedented precision for old
ones, though some of the most relevant constraints, notably $BR(B \to
X_s \gamma)$, still come from the $B$-factories. Besides
the mass splittings and CP violation
in $B^0 - \bar B^0$ and $B_s - \bar B_s$ mixing, 
leptonic, rare semileptonic, and two-body hadronic decays were
discussed at this conference. Important theoretical tools
are lattice QCD and expansions in $\Lambda/m_b$. More specifically,
\begin{itemize}
  \item Lattice QCD, reviewed by J{\"u}ttner \cite{talk-Juettner}, can provide
    $B$ decay constants, comprising the nonperturbative input
    to $B \to \tau \nu$ and $B_q \to \mu^+ \mu^-$ decay,
    for dynamical, light quarks at their physical masses, with uncertainties of a few
    percent.  Matrix elements of $\Delta B=2$ operators needed for
    mixing can also be calculated with uncertainties of order 5\%,
    and the same is true for form
    factors for a decay into a
    stable particle (with regard to QCD) such as a pion or a $K$, as
    long as the momentum transfer $q^2$ is close to maximal (the energy of the light
    particle is small). One needs to treat the $b$-quark in an effective theory (HQET or
    NRQCD), which can however be systematically implemented on  a  lattice. 
    First calculations of $B \to K^*$ form factors
    have also appeared, but only for unphysical quark masses for which
    the $K^*$ is stable. The realistic case of, for instance,
    $B \to K^* [\to K \pi] \mu^+ \mu^-$ would require a $B \to K \pi$ form factor
    calculation, which is at an early stage conceptually
    \cite{Briceno:2014uqa}.
    A complementary method, light-cone sum rules, was discussed by
    Khodjamirian; and can be used to obtain form factors at small
    $q^2$ (as well as other nonperturbative quantities quantities). It
    also has the ability to describe  a $B \to K \pi$ (or similar)
    form factor. Form factor calculations in a quark model were the
    topic of a talk by Hernandez \cite{talk-Hernandez}.
    
  \item Observables related to mixing, the topic of a talk by Bobeth
    \cite{talk-Bobeth}, comprise the mass difference,
    the lifetime difference, time-dependent CP violation in several modes, and CP
    violation in mixing (the flavour-specific or semileptonic CP
    asymmetries). Data show a good overall degree of consistency
    with the SM (Figure \ref{fig:bmix}).
\begin{figure}[h]
\begin{center}
\includegraphics[scale=0.35]{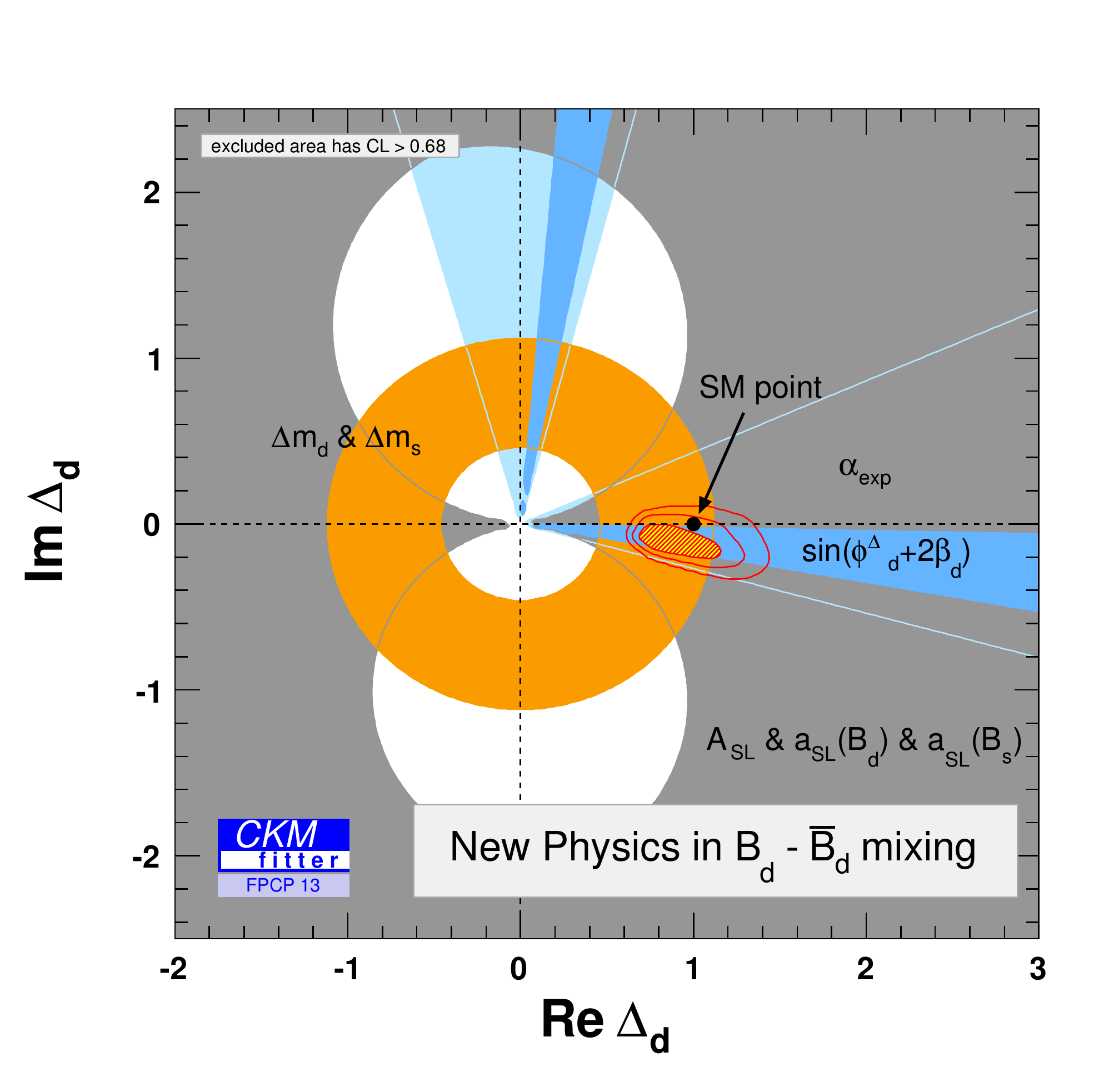}
\hskip1cm
\includegraphics[scale=0.35]{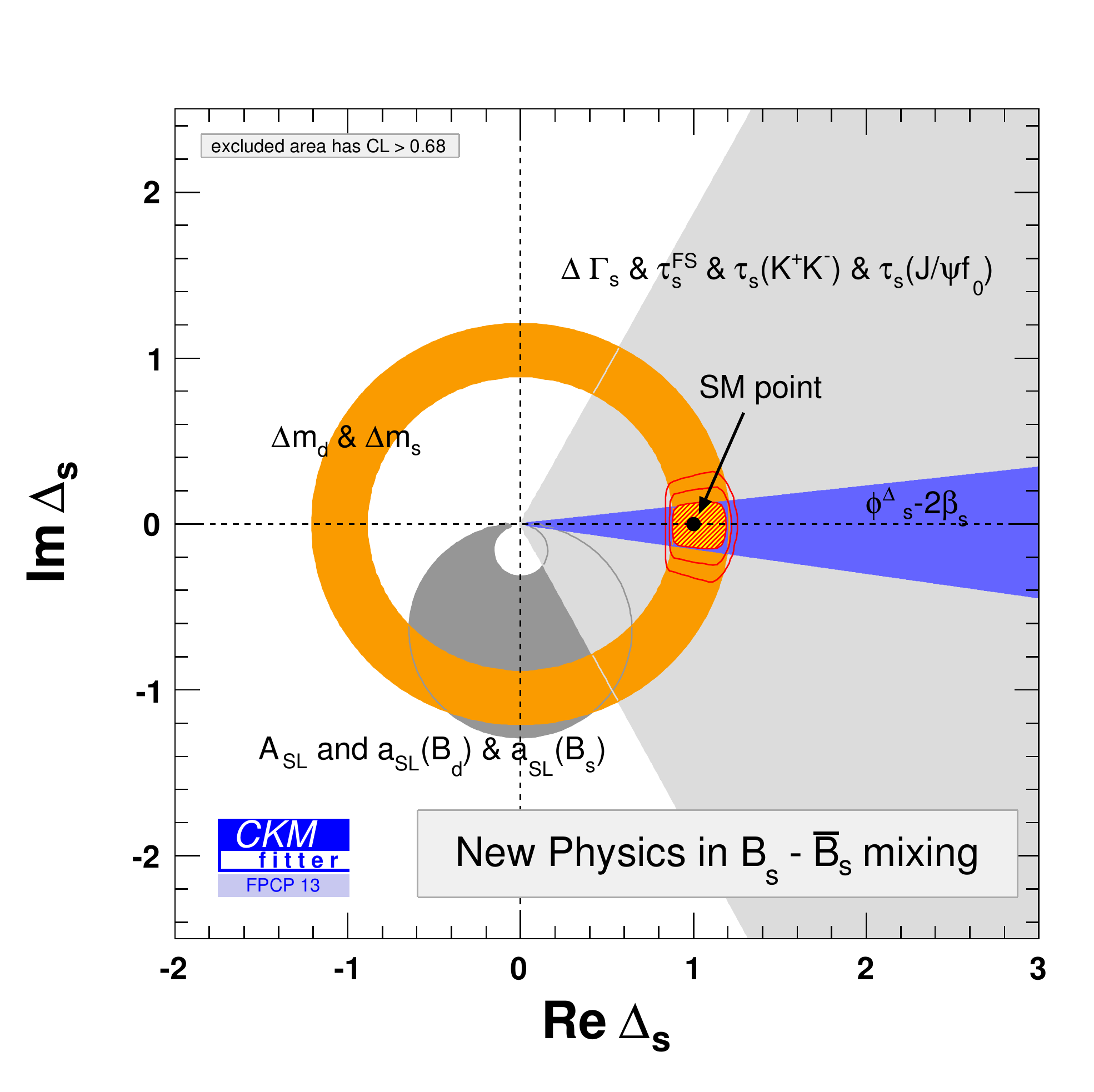}
\end{center}
\caption{\label{fig:bmix} Combined constraints on $B_d-\bar B_d$
  mixing (left) and $B_s-\bar B_s$ mixing (right)  by the CKMfitter
  group (as of FCPC 2013) \cite{Charles:2004jd}.}
\end{figure}
In the case of $B_d$ mixing, there is a mild tension. In the case of
$B_s$ mixing, the best fit is right on top of the SM expectation. One
should note however that the consistency between the different inputs
is not good, in particular, the D0 measurement of
the dimuon charge asymmetry, which depends on both the semileptonic CP
asymmetry and on the lifetime differences, is about $3\sigma$ away from the SM
prediction. Possible interpretations in terms of new decay modes of
$B_d$ were discussed by Bobeth \cite{talk-Bobeth}. In particular, it
appears possible in a model-independent fashion
to explain the discrepancy between the D0 result and
other data with extra contributions to the lifetime difference $\Delta
\Gamma_d$ (as opposed to the semileptonic CP asymmetry) from either
four-quark operators or semileptonic operators with $\tau$ fields,
without getting disagreement with exclusive $B$ decay data.

Another issue that will become important with the increased accuracy with
which LHCb measures time-dependent CP violation in $B_d$ and $B_s$
decays is the ``penguin pollution'' in the various decay modes,
discussed by Fleischer at this conference \cite{talk-Fleischer}
together with $U$-spin-symmetry-based methods for controlling these
effects, which are not yet taken into account in official projected sensitivities. 

Two-body charmless $B$ decays exhibit QCD factorisation in the heavy
quark limit, i.e., they factorise into perturbatively calculable kernels
and nonperturbative objects such as form factors. Bell reported on the
theory status including ongoing computations which will allow to
compute direct CP asymmetries to next-to-leading order \cite{talk-Bell}.

  \item The combined CMS-LHCb results on leptonic decay
    $B_s \to \mu^+ \mu^-$ and $B_d \to \mu^+ \mu^-$  read \cite{Tolk-implications}
$$
  BR(B_s \to \mu^+ \mu^-)^{\rm exp} = 2.8^{+0.8}_{-0.6} \times 10^{-9}, \qquad
  BR(B_d \to \mu^+ \mu^-)^{\rm exp} = 3.9^{+1.6}_{-1.4} \times 10^{-10} ,
$$
and are consistent with SM expectations of
$$
    BR(B_s \to \mu^+ \mu^-) = (3.65 \pm 0.23) \times 10^{-9}, \qquad
    BR(B_d \to \mu^+ \mu^-) = (1.06 \pm 0.09) \times 10^{10} ,
$$
at the $1.2\sigma$ and $2.2\sigma$ levels, respectively, also showing
a certain tension (which is $2.3\sigma$ in the ratio \cite{Tolk-implications}).
I find it extremely impressive that both rates are already measured to
better than 40\% accuracy, in spite of their extreme smallness, which
shows the potential to find even small deviations from the SM, such as
through a modified $Zbs$ vertex. The theoretical expectations given above
are the outcome of an NNLO-QCD and NLO-electroweak calculation,
subject of a talk by Steinhauser \cite{talk-Steinhauser}, which
also investigated issues such as the fate of the helicity suppression
underlying the smallness of the rate under soft photon emission. The theory
errors, which now are below $10 \%$, are dominated by parametric
uncertainties from $f_{B}$ and $f_{B_s}$, which will further reduce
with progress on the lattice.

\item The most intriguing picture presents itself when one goes to rare
semileptonic $b \to s l l$ decays, reviewed by \cite{talk-MartinCamalich}, 
 which entail several anomalies. The bulk of the information is
provided by the angular analysis of $\bar B
\to \bar K^{*} \mu^+ \mu^-$, which involves 12 separate angular
observables, each of which a function of the dilepton invariant mass.
A global fit to coefficients in the BSM effective Lagrangian indicates
a discrepancy with the SM (Figure \ref{fig:anomaly} (left);
see also \cite{Descotes-Genon:2013wba}).
\begin{figure}[h]
\begin{center}
\includegraphics[scale=0.35]{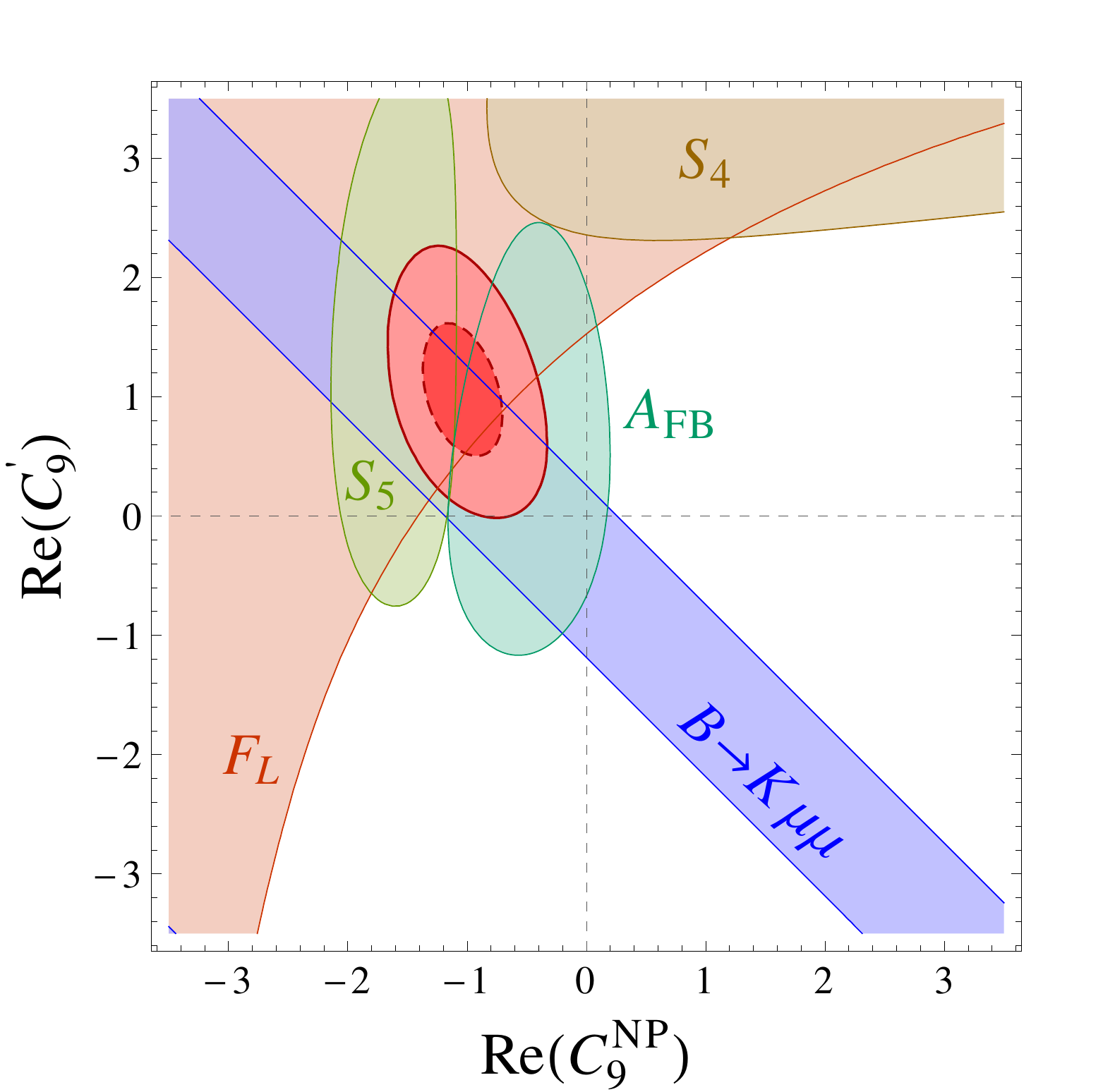}
\hskip1cm
\includegraphics[scale=0.2]{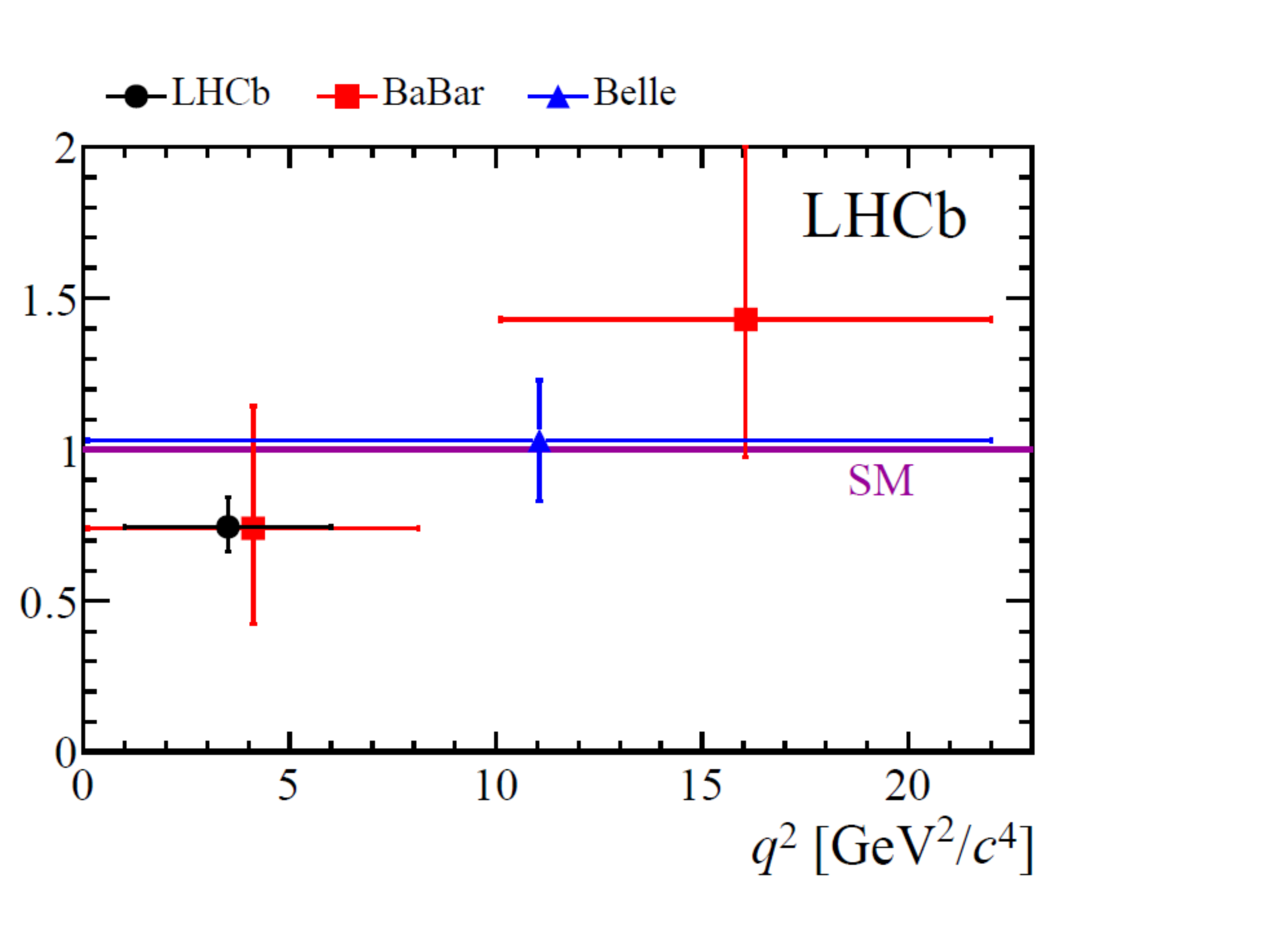}
\end{center}
\caption{Left: Fit of $b \to s l l$ data to two Wilson coefficients in the effective BSM
  hamiltonian \cite{Altmannshofer:2013foa}.
 Right: Lepton-number-(non)universality ratio $R_K = \Gamma(B^+ \to K^+
 \mu^+ \mu^-)/\Gamma(B^+ \to K^+ e^+ e^-)$ \cite{Shires-talk,Aaij:2014ora}.
\label{fig:anomaly}  }
\end{figure}
The effect can be attributed to a sizable negative BSM shift of the Wilson
coefficient $C_9$, multiplying the operator $Q_9 \propto (\bar s_L \gamma^\mu b_L)
(\bar \mu \gamma_\mu \mu)$.
The devil may, however, be in the detail. The angular observables are
quadratic functions in helicity amplitudes, the most complicated of
which are the vector amplitudes, schematically
$$
   H_V(\lambda) \propto \tilde V_\lambda(q^2) C_9 + \frac{2\,m_b
     m_B}{q^2} \tilde T_\lambda(q^2) C_7 - \frac{16\,\pi^2 m_B^2}{q^2}
     h_\lambda(q^2) .
$$
They involve interference of three different terms, including two
nonperturbative local form factors and a further, nonlocal
nonperturbative term. In order to claim an effect of the stated
significance, the necessary precision on the form factors at
small dilepton mass, more specifically on deviations from their
relations in the heavy-quark limit, is on the order of 5\% (Figure
\ref{fig:C9anomaly} left), and in my view it is doubtful that
existing calculations (based mainly on light-cone sum rules) can be
trusted to such a high level of precision. A conservative treatment
of the power corrections \cite{Jager:2012uw} (see also \cite{Beaujean:2013soa})
results in a lower significance of the effect (Figure
\ref{fig:C9anomaly} right), of about $1\sigma$ only.

On the other hand, the picture of a
negative correction to the SM Wilson coefficient $C_9$ is
consistent with the relatively low branching ratio in the decay
$B^+ \to K^+ \ell^+ \ell^-$. Also here the conclusion depends somewhat
on nonperturbative form factor normalisations. The latter largely drop
out of the ratio
$
R_K = \Gamma(B^+ \to K^+ \mu^+ \mu^-)/\Gamma(B^+ \to K^+ e^+ e^-)
= 0.745^{+0.090}_{-0.074} 
$
\cite{Aaij:2014ora} (in the bin $1\, {\rm GeV}^2 < q^2 < 6\, {\rm GeV}^2$),
which points at lepton universality
violation.  If this result stands, it could indeed be accommodated by a
muon-specific negative shift to $C_9$,  consistent with what is
inferred from the $B \to K^* \mu^+ \mu^-$ angular analysis,
but in the muonic operator only (as opposed to that containing the
electronic vector current).
Such scenarios are possible in UV-complete models, 
see e.g.\ \cite{Altmannshofer:2014cfa}.
In this context an interesting new development
first shown at this conference is the systematic use of $SU(2) \times
U(1)_Y$-invariant effective theory, which appears justified by the
likely hierarchy $M_{\rm NP} \gg M_Z$. This rules out tensor operators
as an explanation of any of the anomalies and allows to correlate the $R_K$ measurement
with  $B_s \to \mu^+ \mu^-$ decay, and to rule out any role of scalar
or pseudoscalar operators, leaving only the semileptonic
current-current operators which includes the operator $Q_9$ mentioned above.
\begin{figure}[h]
\begin{center}
\includegraphics[scale=0.35]{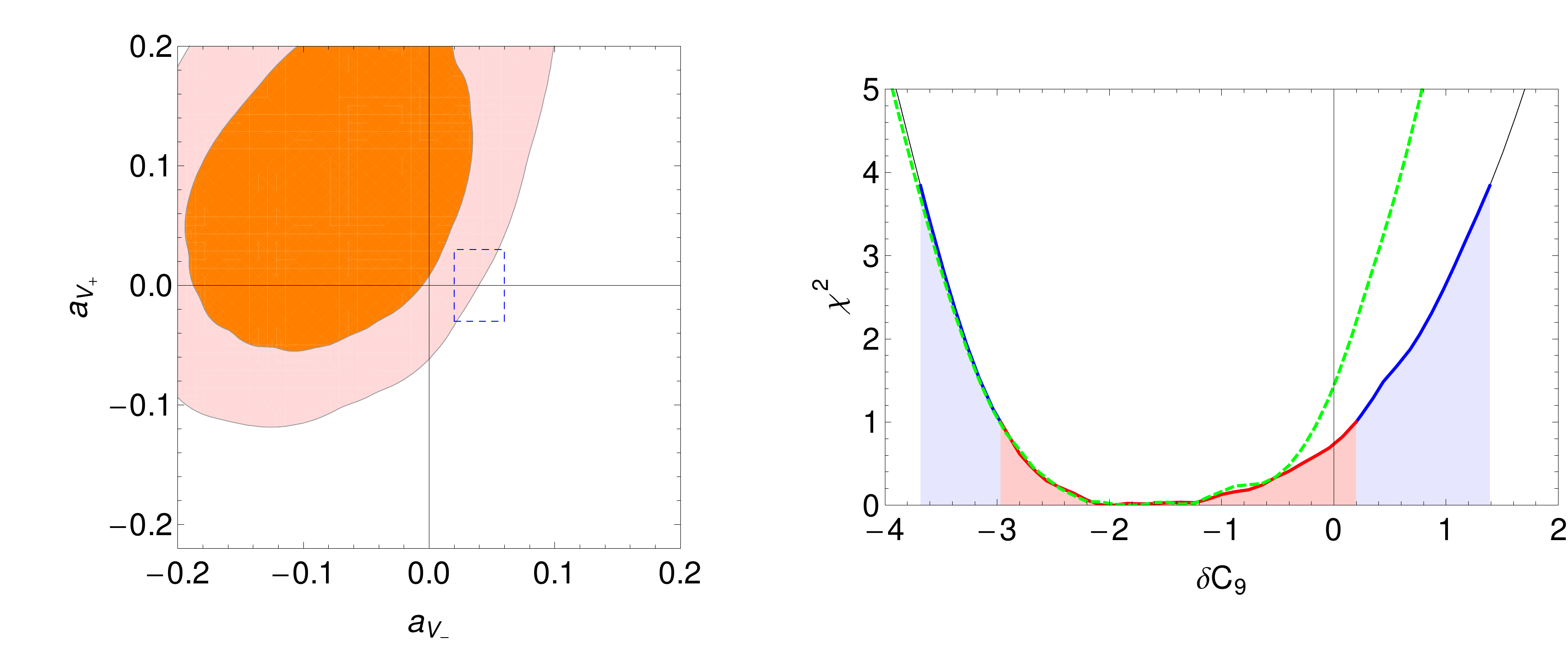}
\end{center}
\caption{\label{fig:C9anomaly}  Left: profile $\chi^2$ of the angular
  observable $P_5'$ in the [1..6] GeV${}^2$ bin, as a function of two
  power correction parameters. Right: profile $\chi^2$ for the Wilson
  coefficient $C_9^{\rm NP}$
  \cite{talk-MartinCamalich,JaegerMartinCamalich}. }
\end{figure}
Finally, the same decay $B^+ \to K^+ \mu^+ \mu^-$ also
shows a rather pronounced resonant structure above the open charm
threshold, the significance of which is unclear at present.

\end{itemize}
The intriguing situation on several fronts makes $B$-physics a
definite place to watch for upcoming developments, such as updates of
the angular analysis in $B \to K^* \mu^+ \mu^-$ with the full run I
dataset, the first angular analysis in $B \to K^* e^+ e^-$, and
high-statistics results based on the upcoming LHC run II and,
before the decade is out, from the Belle 2 super flavour factory.

The top quark was also discussed at this meeting, in
talks by Pecjak \cite{talk-Pecjak} and by Zhang \cite{talk-Zhang}.
Due to the unprecedented top samples at the LHC, it has started to
become the subject of precision studies, among them searches for BSM
effects in flavour-changing top decays. As with the lighter flavours,
assuming a little hierarchy the BSM contributions to
top production and decay can be described in terms of effective field
theory and the data used to put model-independent constraints on its
operator (Wilson) coefficients \cite{talk-Zhang}.

\section{Leptons}

In the lepton sector, flavour is known to be violated by neutrino
oscillations. It is widely expected that lepton number is also
violated, as happens in the seesaw models, which would give rise
to the unique (and lepton-number-violating)
dimension-5 operator beyond the Standard Model.
All data is presently consistent with this picture, with exception of
the well-known LSND anomaly. Status
and prospects were reviewed by Pakvasa
\cite{talk-Pakvasa} and by Palazzo \cite{talk-Palazzo}. After the
establishment of $\theta_{13} \approx 9^\circ \not= 0$, the main
questions are:
(i) do neutrino masses have a lepton-number-violating origin
(search for neutrinoless double beta decay),
(ii) is CP violated in the lepton/neutrino sector, (iii) is the neutrino mass
hierarchy normal or inverted, (iv) are there sterile neutrinos
(and if so at what scale)? There are indications in the global fit
for a nonzero (``Dirac'', lepton-flavour-violating)
CP-violating phase $\delta \in (\pi, 2\pi)$, although the significance
depends on the mass hierarchy and is never above $2\sigma$.
This and the other questions will, however, take a long time to resolve
definitely. If lepton number is indeed violated, one can also have CP
violation through the ``Majorana'' phases in the neutrino mass matrix.

The fact that lepton flavour violation is observed implies that
charged lepton-flavour-violating processes must also occur,
such as $\mu \to e \gamma$. They were the subject of a talk
by Paradisi \cite{talk-Paradisi}. In the
SM (supplemented by neutrino masses),
the tiny masses of the neutrinos (when compared to the $W$ mass)
imply a near-perfect GIM
cancellation, with no hope to see a signal. The situation can be very
different beyond the SM. SUSY seesaw models, for example, communicate the
flavour violation above the seesaw scale
through to low-energy scales via renormalisable terms,
such as off-diagonal elements of the slepton  mass matrices. The
resultant SUSY contributions can
saturate the current bounds on many lepton-flavour-violating decay
modes. An interesting aspect are relations within the general MSSM between
the radiative lepton-flavour-violating  decays and the
flavour-conserving but BSM-sensitive anomalous magnetic moment
$(g-2)_\mu$ of the muon. The interest derives from the fact that the
latter has shown a $3\sigma$ deviation from the SM prediction for a
number of years now. The correlation and the theory of $(g-2)_\mu$
was reviewed in more detail by Velasco-Sevilla \cite{talk-Velasco}.

\section{Origin of matter}

The most important aspect of CP violation may be that our existence
may be thanks to it. More specifically, the observed excess of matter
over antimatter in the universe can be explained by out-of-equilibrium
CP-violating processes (together with violation of baryon number and charge
conjugation symmetry) in the early universe.
There are many models and mechanisms providing
a successful baryogenesis, and it is impossible to conclude from the
observation of the baryon asymmetry alone which one is correct.
Some of the more compelling ones are, however, connected
to CP-violating flavour physics at the TeV scale. Electroweak
baryogenesis in a Two-Higgs-doublet model was discussed by Huber
\cite{talk-Huber}. Here new degrees of freedom near the weak scale
effect a strong first-order electroweak phase transition; the CP
violating processes occuring in the course of this are found to be
correlated with rare $B$ decay and $B_s - \bar B_s$ mixing, as well as
electric dipole moments of elementary particles.
Schwaller discussed leptogenesis \cite{talk-Schwaller}, whereby heavy SM singlets in a
seesaw model decay to Higgs + lepton in a CP-asymmetric fashion; the
necessary baryon number violation (which converts part of the lepton
number into baryon number) takes place at much lower
energies/temperatures via sphaleron processes. An important topic are
thermal corrections to CP violation and a proper account of the tau
Yukawa coupling. 

\section{Outlook}
\begin{figure}[h]
\begin{center}
\includegraphics[scale=0.5,trim=0 0 0 0]{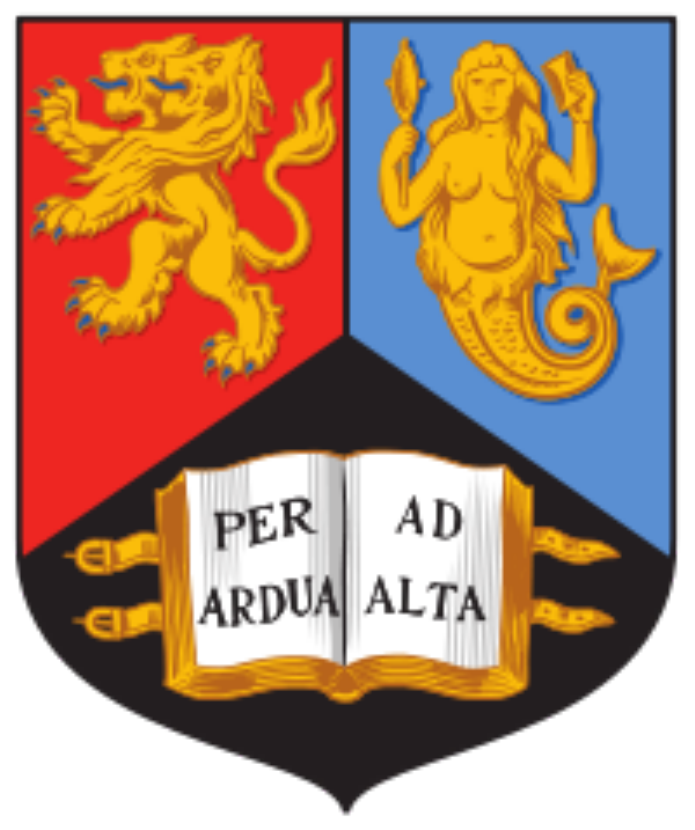}
\hskip2cm
%
\includegraphics[scale=0.6]{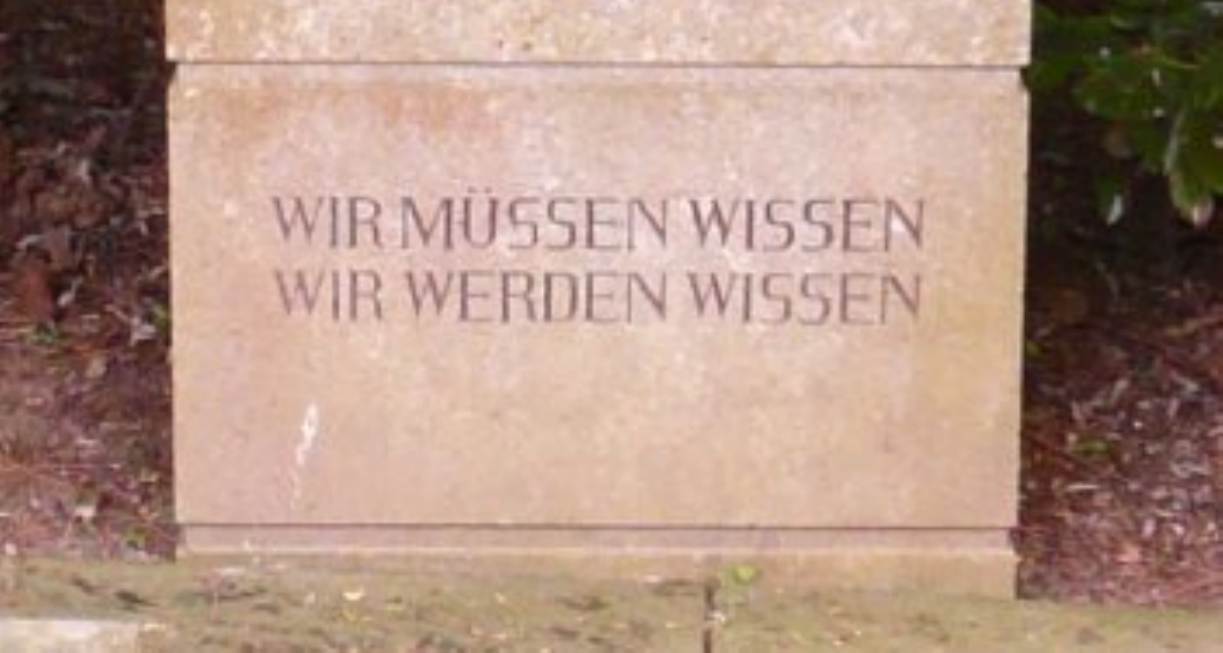}
\end{center}
\caption{\label{fig:motto} Left: The University of Birmingham's motto,
  ``Through struggle to high energies.'' Right: Hilbert's tomb
  [Source: Wikimedia Commons].} 
\end{figure}

In conclusion, while it can be at times disappointing that the eagerly
anticipated signatures of new physics below the TeV scale have not
materialised, we are in the lucky circumstance of being showered with
new results on a regular basis. While, for the time being, most of them
imply ever tighter constraints on BSM physics, we should note that
the arguments motivating physics at or near the TeV have not
diminished.
Any of the several tensions mentioned can, with more data and/or better
theory, turn into significant falsifications of the Standard Model;
and they may in fact already be pointing at a consistent picture.
That the search for new physics would be a laborious affair
was, by the way, predicted here in Bimingham, as the
university motto attests (Figure \ref{fig:motto} left). Thanks to our
experimental colleagues, there is no shortage of work for flavour
theorists these days. While, as in any ambitious project,
success is not certain, I am confident that, in the end, Hilbert will be
proven right (Figure \ref{fig:motto} right).

\section*{Acknowledgment}
I would like to thank the organisers warmly for such an exciting and perfectly
organised conference. I am grateful to the many theory speakers
for their clear presenations,
which made my task easier than it could
have been, and to Alexey Petrov for photographs used on my slides
(not reproduced here). Support from the UK STFC under
grants ST/J000477/1 and ST/L000504/1 is acknowledged.

\section*{References}

\bibliography{beach}
\bibliographystyle{iopart-num}

\end{document}